\documentclass{article}
\usepackage[normalem]{ulem}
\usepackage{amsthm}
\usepackage{mathtools}
\usepackage{thmtools}
\declaretheoremstyle[headfont=\normalfont]{normalhead}
\usepackage{color}
\usepackage{graphicx}
\usepackage{morefloats}
\usepackage{hyperref}
\usepackage{ amssymb }
\usepackage{textcomp}
\usepackage{url}
\usepackage{float}
\usepackage{biblatex}
\newtheoremstyle{mydef}
{\topsep}{\topsep}%
{}{}%
{\itshape}{}
{\newline}
{%
  \rule{\textwidth}{0.0pt}\\*%
  \thmname{#1}~\thmnumber{#2}\thmnote{\-\ #3}.\\*[-1.5ex]%
  \rule{\textwidth}{0.0pt}}%
  
\begin{document}

\theoremstyle{mydef}
\newtheorem{conjecture}{Conjecture}
\newtheorem{theorem}{Theorem}
\newtheorem{question}{Question}
\newtheorem{remark}{Remark}
\newtheorem{proposal}{Proposal}
\newtheorem{lemma}{Lemma}
\newtheorem{corollary}{Corollary}
\newtheorem{observation}{Observation}
\author{Aditya Mittal and Karthik Mittal}

\date{February 19, 2021}

\title{Determining Smallest Path Size of Multiplication Transducers Without a Restricted Digit Set}
\maketitle
\begin{abstract}
\hskip -.2in
\noindent

Directed multiplication transducers are a tool for performing non-decimal base multiplication without an additional conversion to base 10. This allows for faster computation and provides easier visualization depending on the problem at hand. By building these multiplication transducers computationally, new patterns can be identified as these transducers can be built with much larger bases and multipliers. Through a Python-based recursive approach, we created artificial multiplication transducers, allowing for the formation of several unique conjectures specifically focused on the smallest closed loop around a multiplication transducer starting and ending at zero. We show a general recursive pattern for this loop; through this recurrence relation, the length of the smallest closed loop for a particular transducer base \textit{b} along with the range of multipliers having this particular length for multiplier \textit{m} was also identified. This research is expected to be explored further by testing reductions of the digit set and determining whether similar properties will hold.

\end{abstract}
\vspace{\baselineskip}
\section{Introduction}

 \hspace*{5mm} Directed multiplication transducers are tools to multiply a number by base $b$ without the need for conversion into an intermediary base such as base 10. These transducers can run at computationally faster speeds due to this property. Finding patterns in these transducers (e.g. recursive formulas and minimum path lengths for specific base and multiplier transducers) can introduce faster methods for finding bases and create new breakthroughs in the field of quotient sets.   \vspace{\baselineskip}

 This paper analyzes multiplication transducers with no excluded digits and determines patterns in paths (where all states in the path are distinct) within the multiplication transducer starting and ending with zero as a state. For all bases $b$ and multipliers $m$, a generalized conclusion can be made on the length and values of states within the shortest path. \vspace{\baselineskip}
 
 This paper presents a formula for the length of the shortest possible quotient set given base $b$ and multiplier $m$ and provides a recursive solution for finding this using depth first search and Python libraries. This newfound method of identifying these paths can lead to faster computational analysis regarding the multiplication of numbers in different bases; research will be performed in the future to find this formula for restricted digits.  \vspace{\baselineskip}
 
 The applications of multiplication transducers are varied, but they can be seen mostly in number theory and automata. By recognizing these paths, faster computations can be made, and shortcuts can be found to emerging problems within the field.

\subsection{Overview of Multiplication Transducers}
\hspace*{5mm}Multiplication transducers can consolidate the information stored in base multiplication into an easily explainable diagram \cite{Blanchard1992} \cite{sisneros-thiry}. Base multiplication has five essential components when in step $i$: the carry-in value $c_i$, the read value $r_i$, the total value $t_i$, the write value $w_i$ and the carry over value $c_{i+1}$. Elementary one-digit base ten multiplication (which has only one step) can be discussed in order to understand these characteristics.  \vspace{\baselineskip}

\noindent
\textbf{Example:} Let's take a base ten example where we are multiplying 5 by 6. In variables, this means $m = 6$, $b = 10$, and $r = 5$. We know this can be done with multiplication, but this can also be completed using multiplication transducers. Some notes can be taken:
\begin{itemize}
    \item The carry-over value for the first step is $0$, because we aren't carrying over anything. This means that $c_0 = 0$.
    \item The read value for the first step is $5$, which means $r_0 = 5$. Note that if the read value was $15$ for instance, then $r_0 = 5$ and $r_1 = 1$. 
\end{itemize}

\noindent 
\textbf{Step 1} ($i=0$): $c_0 = 0$ (our initial state) and $r_0 = 5$. The following is determined: 
\begin{itemize}
  \item We first compute the total. This can be done by multiplying our current read by our multiplier, and then adding over any carry values from earlier. This gives us a total of $(5*6) + 0 = 30$.
  \item Next, we have to calculate the carry value. Since $30$ consists of $3 \, 10$'s, this means that the carry value or $c_1 = \lfloor{\frac{30}{10}}\rfloor = 3$. 
  \item Lastly, we have to calculate the write value or the value that we will write down from our first step. This is calculated by finding the remainder when dividing the total from the base. This means that the first write value or $w_0 = 30 \, (\text{mod} \, 10) = 0$.
\end{itemize} \vspace{\baselineskip}

\noindent
Some things to note before the next step:
\begin{itemize}
    \item The carry-over value for the second step is 3, which means that $c_1 = 3$.
    \item The read value for the second step is 0 because our read was only one digit. This means that $r_1 = 0$.
\end{itemize}

\noindent 
\textbf{Step 2} ($i=1$): $c_1 = 3$ and $r_0 = 0$. The following is determined: 
\begin{itemize}
  \item We first compute the total. This can be done by multiplying our current read by our multiplier, and then adding over any carry values from earlier. This gives us a total of $(6*0) + 3 = 3$.
  \item Next, we have to calculate the carry value. Since $3$ consists of $0 \, 10$'s, this means that the carry value or $c_1 = \lfloor{\frac{3}{10}}\rfloor = 0$. 
  \item Lastly, we have to calculate the write value or the value that we will write down from our first step. This is calculated by finding the remainder when dividing the total from the base. This means that the first write value or $w_0 = 3 \, (\text{mod} \, 10) = 3$.
\end{itemize} \vspace{\baselineskip}

Note that both our read value and carry value for the third step are zero. This means that our total will be zero, which means that the write value will be zero. Therefore, we are done with our calculations. We have $w_0 = 0$ and $w_1 = 3$, so our final answer is $w = 30$. \vspace{\baselineskip}

This can now be expressed with variables. For a one-digit $r$ by $m$ base ten multiplication operation, $c_0 = 0$ since there is no underlying carry value from a previous multiplication. $r$ is the read value while $m$ is the multiplier. The total can be represented as: \begin{equation} t_0 = r_0m + c_0 = rm + c_0 = 10c_1 + w_0\end{equation} 

 The equation above is justified since the carry value $c_{i+1}$ is always written when the write value is too large to be expressed, like in base ten multiplication operations. In this case, $c_1$ would be $w_1$ since $r_1 = 0$ and $r_1m = 0$ (assuming that $r$ and $m$ are both one digit). However, when dealing with larger numbers, this number will be carried over to the next multiplication, until the operation is solved, so that the carry over value $c_{i+1}$ becomes the carry in value. \vspace{\baselineskip}
 
 We can expand this concept to any $r$ by $m$ multiplication operation in base $b$. Let $l_w$ be the length of the final write value $w$ in base $b$ and $l_r$ be the length of the read value in base $b$. Then, for $r$ and $w$ in base ten,
 \begin{equation}
     w = \sum_{i=0}^{l_w - 1} w_ib^{i}, r = \sum_{i=0}^{l_r - 1} r_ib^{i}
 \end{equation} For $r$ and $w$ in base $b$, $r = [r_{l_r-1}r_{l_r-2}...r_0]_b$ and $w = [w_{l_w-1}w_{l_w-1}...w_0]_b$. The read value $r_i$ for multi-digit operations will be the last value index of $r$ for the first operation ($i$ = $0$), the penultimate value index for the second ($i$ = $1$), and so on. Therefore, the generalized total for step $i$ in a base $b$ multiplication operation can be written as: \begin{equation} t_i = r_im + c_i = bc_{i+1} + w_i\end{equation}
 
 The write value $w_i$ and the carry over value $c_{i+1}$ are found with: \begin{equation} c_{i+1} = \left \lfloor{\frac{t_i}{b}} \right \rfloor \text{; } w_i = t_i \text{ (mod b)} \end{equation}
 
\begin{figure}
  \centering
  \includegraphics[width=0.5\textwidth,width=0.4\textheight, trim = 0cm 0.5cm 0cm 0.7cm, clip]{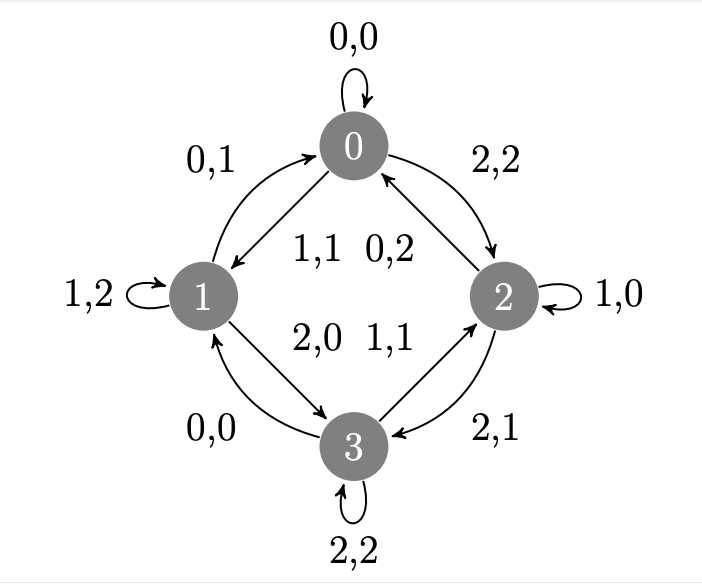}
  \textbf{\caption{
    \label{transducer}
    Representation of the multiplication transducer $T_{4,3}$.
  }}
\end{figure}

 A multiplication transducer represents all the distinct combinations of this base multiplication for a predefined multiplier $m$ and base $b$, iterating through the different possible combinations between the read value $r$ and the carry in value $c_{i+1}$. In multiplication transducers, carry in values are often represented by states (denoted by circles). As shown in Figure \ref{transducer}, the corresponding read value $r_i$ and write value $w_i$ is written adjacent to the arrow pointing from state $c_i$ to $c_j$, in the notation ($r_i$, $w_i$). The total is calculated using a different interpretation of the equation above: \begin{equation} t_i = r_{i}m + c_{i} = bc_{j} + w_{i}\end{equation}
 
 The carry value $c_i = \{0, 1, ..., m-1\}$ and read value $r_i = \{0, 1, ..., b-1\}$ represent the total of $b^{2}$ combinations in the multiplication transducer of designated base $b$ and multiplier $m$ or $T_{m,b}$. \vspace{\baselineskip} 

\subsection{Base 3 Example}
\hspace*{5mm} Let's take an example of $m = 4$ and $b = 3$. Let $r = [20]_{10} = [202]_3$. We can use multiplication transducers to determine $[202]_3 * 4$ without converting to base $10$. \vspace{\baselineskip}

\noindent 
\textbf{Step 1} ($i=0$): $c_0 = 0$ (our initial state) and $r_0 = 2$. The following is determined: 
\begin{itemize}
  \item $t_0 = r_0m + c_0 = 2*4 + 0 = 8$
  \item $c_1 = \lfloor{\frac{t_0}{b}} \rfloor = \lfloor{\frac{8}{3}} \rfloor = 2$
  \item $w_0 = t_0 \text{ (mod b)} = 8 \text{ (mod 3) } = 2$
  \item $c_0$ to $c_1$: The state in Figure \ref{transducer} changes from $0$ to $2$ with a read value of $2$ and a write value of $2$. This corresponds with our calculations in Step 1, as $r_0 = 2$ and $w_0 = 2$.
\end{itemize} \vspace{\baselineskip}

\noindent
\textbf{Step 2} ($i = 1$): $c_1 = 2$ and $r_1 = 0$. The following is determined:
\begin{itemize}
  \item $t_1 = r_1m + c_1 = 0*4 + 2 = 2$
  \item $c_2 = \lfloor{\frac{t_1}{b}} \rfloor = \lfloor{\frac{2}{3}} \rfloor = 0$
  \item $w_1 = t_1 \text{ (mod b)} = 2 \text{ (mod 3) } = 2$
  \item $c_1$ to $c_2$: The state in Figure \ref{transducer} changes from $2$ to $0$ with a read value of $0$ and a write value of $2$. This corresponds with our calculations in Step 2, as $r_1 = 0$ and $w_1 = 2$.
\end{itemize} \vspace{\baselineskip}

\noindent
\textbf{Step 3} ($i = 2$): $c_2 = 0$ and $r_2 = 2$. The following is determined: 
\begin{itemize}
  \item $t_2 = r_2m + c_2 = 2*4 + 0 = 8$
  \item $c_3 = \lfloor{\frac{t_2}{b}} \rfloor = \lfloor{\frac{8}{3}} \rfloor = 2$
  \item $w_2 = t_2 \text{ (mod b)} = 8 \text{ (mod 3) } = 2$
  \item $c_2$ to $c_3$: The state in Figure \ref{transducer} changes from $0$ to $2$ with a read value of $2$ and a write value of $2$. This corresponds with our calculations in Step 3, as $r_2 = 2$ and $w_2 = 2$.
\end{itemize} \vspace{\baselineskip}

\noindent
\textbf{Step 4} ($i = 3$): $c_3 = 2$ and $r_3 = 0$ (because $[0202]_3 = [202]_3$). The following is determined:
\begin{itemize}
  \item $t_3 = r_3m + c_3 = 0*4 + 2 = 2$
  \item $c_4 = \lfloor{\frac{t_3}{b}} \rfloor = \lfloor{\frac{2}{3}} \rfloor = 0$
  \item $w_3 = t_3 \text{ (mod b)} = 2 \text{ (mod 3) } = 2$
  \item $c_3$ to $c_4$: The state in Figure \ref{transducer} changes from $2$ to $0$ with a read value of $0$ and a write value of $2$. This corresponds with our calculations in Step 4, as $r_3 = 0$ and $w_3 = 2$.
\end{itemize} \vspace{\baselineskip}

The iteration is terminated when there are no more read values ($i > l_r$) and when $c_{i+1} = 0$. Adding all of the write values will give our final value. We know $w$ in base 3 is $[w_3w_2w_1w_0]_3 = [2222]_3$. Using Equation 2 for $w$, $w = (2 * 3^0) + (2*3^1) + (2*3^2) + (2*3^3) = 80$. We can see this equates to the more familiar base $10$ multiplication of $r * m = 20 * 4 = 80$. \vspace{\baselineskip}

\section{Methods}

The methods below will primarily cover the steps towards finding the minimum length for a path of $c$'s starting and ending at zero for $T_{m,b}$. Finding these minimum lengths (if the length is greater than zero) will allow us to find whether a quotient set exists for $T_{m,b}$; these quotient sets will form the basis of many conjectures that will be outlined later in this paper. 

\subsection{Visualizer}

\hspace*{5mm} The visualizer works by producing an artificial multiplication transducer (with values stored in a data structure) and then transforming this structural representation into a visual one similar to that of Figure \ref{transducer}. A visual representation of the multiplication transducer was generated using Python libraries like Matplotlib in order to further prove the logic behind the conjectures outlined in this paper; however, these visualizers became difficult to read as $b$ and $m$ increase due to the fast growth of combinations between different states. Since the visualizer proves these conjectures for smaller bases/multipliers, this limitation does not hinder how the conjectures are proven. \vspace{\baselineskip}

For making the artificial representation of the transducer, every combination of $c_i$ and $r_i$ was iterated through, and the corresponding $c_{i+1}$ and $w_i$ were calculated from these. Note that Equations 3 and 4 can be used to calculate these values. Since $c_i$ has a possible $m$ values while $r_i$ has $l_r$ values, the runtime complexity of this step is $O(ml_r)$. Note that since there are $m$ carry values, this means that even if $l_r > m$, there are only $m$ calculations that need to be made. Therefore, the runtime complexity for this is $O(m^2)$. \vspace{\baselineskip}

Secondly, Matplotlib was used to convert this data into an visual multiplication transducer. Various functions were used, such as plt.Circle (which created the structure to house the carry-in values), plt.arrow (which created lines between the carry-in values), and plt.text (which helped create text on the graph that made the visualizer easier to view). \vspace{\baselineskip}

\begin{figure}
  \centering
  \includegraphics[width=0.64\textwidth, height=0.4\textheight, trim = 0cm 0cm 0cm 0.6cm, clip]{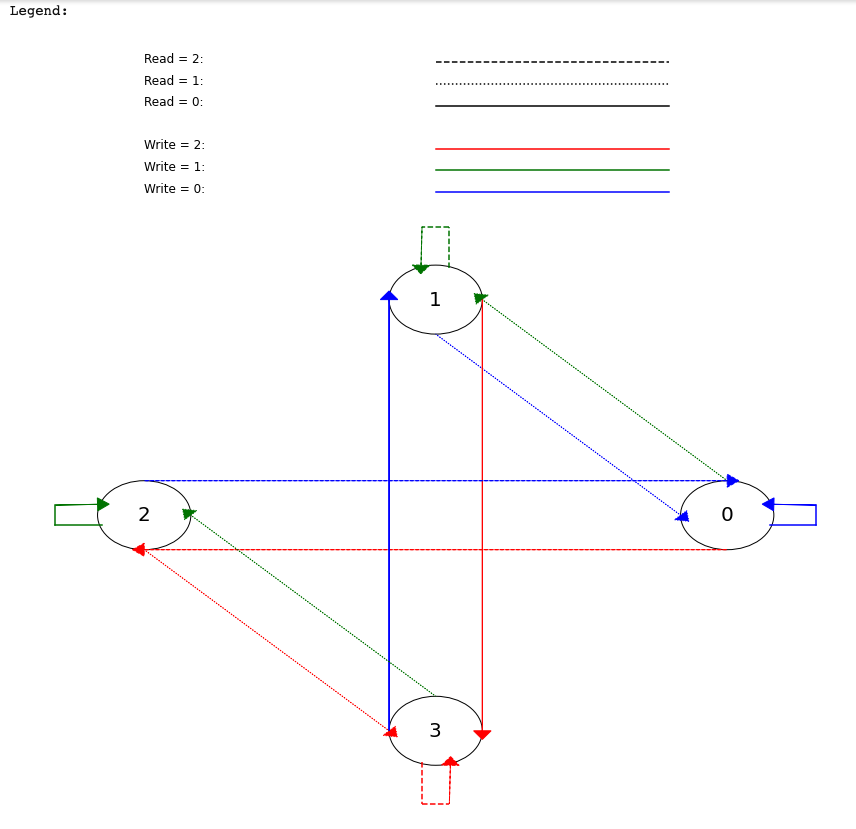}
  \textbf{\caption{
    \label{visualizer}
    Representation of the multiplication transducer $T_{4,3}$ using the visualizer.
  }}
\end{figure}

In addition, different colors and line styles (e.g. blue dashed lines) were used to represent the write and read values respectively for the arrow between the carry-in values. Using these processes, a figure similar to Figure \ref{visualizer} was created. \vspace{\baselineskip}

These particular linestyles were chosen due to their distinctness from the other linestyles that would be represented on the transducer; in order for easier comprehension of the transducer itself, it was necessary to choose differentiating characteristics for each of the linestyles so that the reader can understand which line corresponds a specific read value. \vspace{\baselineskip}

Besides from the simple linestyles provided by Matplotlib (e.g. 'solid', 'dotted', and 'dashed'), more complex linestyles were taken to increase the amount of read values that can be represented on the visualizer; this control was achieved by providing a dash tuple with the form (offset, (on\_off\_seq)). For example, (0, (2, 7, 1, 14)) represents a 2 pt line, a 7 pt space, a 1 pt line, and a 14 pt space with no offset. \vspace{\baselineskip}

A similar process was done for the colors with the write values where only extremely distinct colors were chosen (e.g. red, green, blue, etc.). More nuanced colors were discarded due to their difficult visibility. \vspace{\baselineskip}

\begin{figure}
  \centering
  \includegraphics[width=0.72\textwidth, height=0.45\textheight, trim = 0cm 0cm 0cm 0.2cm, clip]{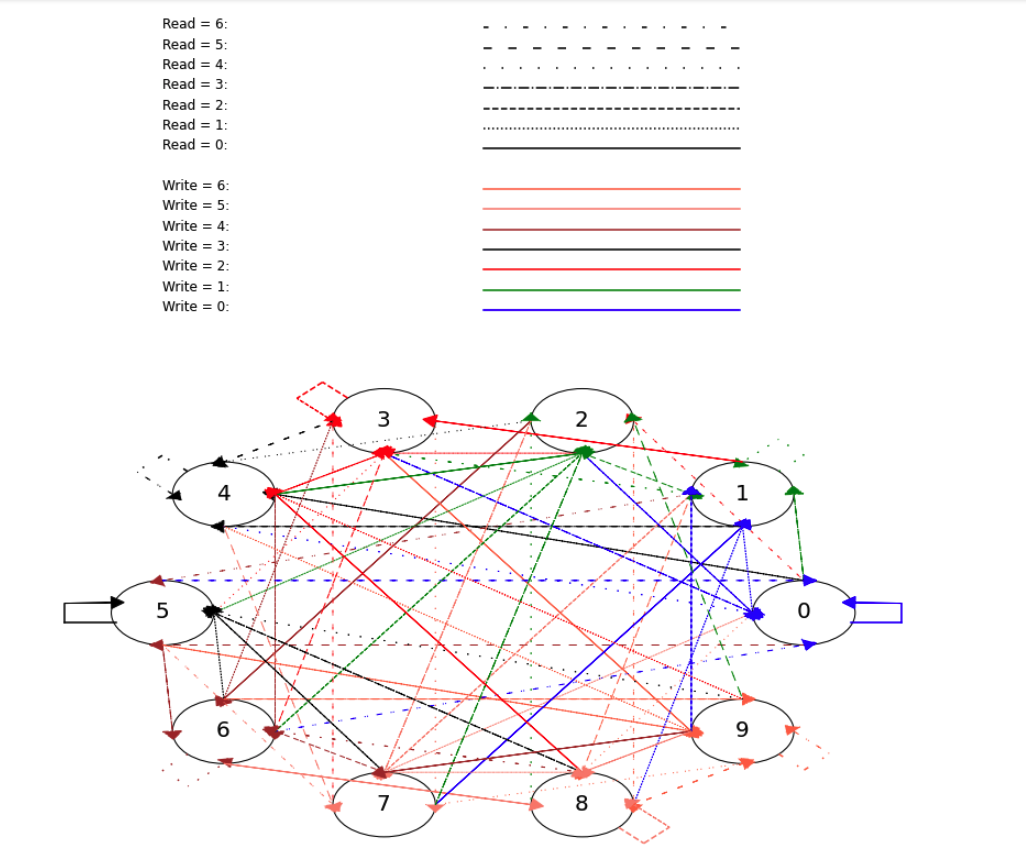}
  \textbf{\caption{
    \label{complex_visualizer}
    Representation of a more complex multiplication transducer $T_{10,7}$ using the visualizer.
  }}
\end{figure}

Note that the visual multiplication transducer should only be used for $m < 8$ and $b < 8$, because values greater than that will produce a multiplication transducer that will be difficult to comprehend (see Figure \ref{complex_visualizer}). The implementation behind the visualizer can be referenced \href{https://github.com/karthikm15/Multiplication-Transducer/blob/main/Matplotlib\%20Visualization.ipynb}{\color{blue}here.}

\subsection{Multiplication Transducer Traversal}

\hspace*{5mm}To find and validate these conjectures, an artificial multiplication transducer was formed in Python by iterating through the possible read and carry value combinations for a particular base $b$ and multiplier $m$. Since multiplication transducers for different bases and multipliers can be difficult to compute and draw non-computationally, a multiplication transducer formed by a computational algorithm provided the perfect method to scale bases and multipliers efficiently. \vspace{\baselineskip}

Two different approaches were mainly used to generate the pathways to make the multiplication transducer itself and to traverse these pathways to find the minimum length path $p$ that starts and ends at zero: the networkx library and the depth first search algorithm. Note that these two different methods were both implemented in order to ensure the validity of the findings outlined in this paper.  \vspace{\baselineskip}

\subsubsection{NetworkX Library}
\hspace*{5mm} The networkx library is a library that provides a platform for reviewing graphs and networks using Python. By building and manipulating complex structures, the networkx library is widely used by computational mathematicians wanting to solve new conjectures in the fields of graph theory. As shown in Figure \ref{transducer}, multiplication transducers can be seen as these networks that networkx manipulates. The states can be interpreted as the vertices of the network while the corresponding arrow with the read and write values represent the edges of the network. \vspace{\baselineskip}

These transducers can be formed computationally using the networkx library through the .add\_node(), and .add\_edge() commands. Since the networkx library has a number of different standard graph algorithms for different niche cases and provides different measures for analysis, it was the optimal library for building a multiplication transducer. Specifically, the .DiGraph() command was used as the multiplication transducer acts as a directed graph since there are arrows pointing to the next possible state inside the transducer (meaning that it has a direction).

After the generation of this multiplication transducer, this directed graph was then traversed using the networkx library to find the shortest possible paths in the graph that start at state zero and end at state zero (one of the conditions necessary for a path to be part of the quotient set). The implementation behind the networkx library can be referenced \href{https://github.com/karthikm15/Multiplication-Transducer/blob/main/networkx.py}{\color{blue} here}. \cite{Everitt2015}

\subsubsection{Depth First Search}

\hspace*{5mm} The depth first search algorithm (DFS) traverses tree structures by starting at the root node and travelling down each possible path to minimize a specified parameter. Although algorithms such as breadth-first search (BFS) have similar time complexities of $O(|V| + |E|)$ with $V$ being the number of states and $E$ being the number of arrows between the states, DFS is more suitable due to its inherent algorithmic structure since the first states explored (e.g. state 0, state 1, etc.) often provide the optimal solution and there are much more solutions farther away from the source. \vspace{\baselineskip}

The generation of the multiplication transducer uses a similar strategy to the one seen in Section 2.1 since the networkx library is not being utilized for graph traversal in this scenario like in Section 2.2.1. \vspace{\baselineskip}

Note that the logic in Section 2.2.1 shown to prove a multiplication transducer to be a directed graph can be used to allow for DFS to be implemented on the artificial transducer. Since this algorithm was run in Python, arrays were used instead of stacks (which follow a last in first out pattern). A recursive formula was mainly utilized to perform this depth first search; the code used can be referenced \href{https://github.com/karthikm15/Multiplication-Transducer/blob/main/DFS.py}{\color{blue} here}. 

 \begin{figure}
  \centering
  \includegraphics[width=0.6\textwidth,width=0.6\textheight, trim = 0cm 0.5cm 0cm 0.7cm, clip]{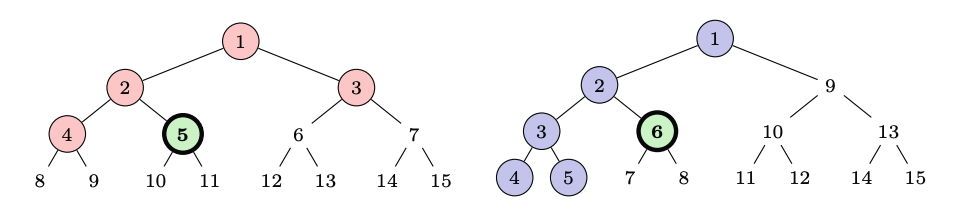}
  \textbf{\caption{
    \label{transducer2}
        Side by side representation of a BFS vs. DFS approach.
  }}
  \small \vspace{\baselineskip}
  These two tree traversals represent the inherent differences behind the BFS and the DFS approach. The DFS (the graph shown on the right) works better for our project as it goes through the foremost nodes first before traversing down the rest of the tree. A DFS implementation was taken since most of the smallest paths will lie in the first section of the transducer itself.
\end{figure}

\subsection{Challenges}
\subsubsection{C++ Implementation}
For most programs, C++ is computationally faster at running algorithms compared to Python; this is why it is often the preferred language for time-intensive operating programs. Therefore, we attempted a C++ implementation for forming the artificial multiplication transducer to reduce the time and space complexity of our operations.

However, the operations done by this algorithm worked slower instead of faster as allocating space to a vector took a computationally intensive time, especially for larger bases and multipliers. Therefore, at the end, we took a Python-based approach to build the transducer. The implementation for building the transducer using C++ can be seen \href{https://github.com/karthikm15/Multiplication-Transducer/blob/main/C\%2B\%2B\%20Implementation.cpp}{\color{blue} here}.

\subsubsection{NetworkX Visualizer}
\hspace*{5mm} We considered the networkx library when building the visualizer to build an efficient and visually appealing model compared to the Matplotlib library. Due to its versatility, we believed that the networkx library could be used to not only build the artificial multiplication transducer but also create it visually; this would provide a consolidated approach for building these multiplication transducers, only involving the use of one library. \vspace{\baselineskip}
 
However, when creating this visualization, there was no option to produce different linestyles or colors, which meant that different read and write values could not be differentiated between. Additionally, the scale of the graph could not be altered, which meant that the multiplication transducer was becoming too cluttered even for small bases. After seeing this effect with networkx, we decided that a manually-made multiplication transducer would be optimal.

\section{Results}

\textbf{Conjecture 1:} For all natural numbers $b$, $m > 1$, the path of carry values $c_i$'s that is the smallest closed loop across a multiplication transducer $T_{m,b}$ starting and ending from $0$ is:
\begin{itemize}
  \item $c_0 = 0$.
  \item $c_1 = \lfloor{\frac{m}{b}}\rfloor$.
  \item $c_i = \lfloor{\frac{c_{i-1}}{b}}\rfloor$ for $i \geq 2$.
\end{itemize}
\textit{Note.} The conjecture has been computationally checked until $b < 2000$ and $m < 2000$. \vspace{\baselineskip}

\noindent
Additionally, note that for the conditions stated in Theorem 1, $c_0 = 0$, so $t_0 = r_0m$ and $c_1 = \lfloor{\frac{r_0m}{b}}\rfloor$. This indicates that $r_0 = 1$, since $c_1 = \lfloor{\frac{m}{b}}\rfloor$ as stated in the theorem. Similarly, since $p_i = \lfloor{\frac{p_{i-1}}{b}}\rfloor$, $c_2 = \lfloor{\frac{p_1}{b}}\rfloor = \lfloor{\frac{c_1}{b}}\rfloor = \lfloor{\frac{t_1}{b}}\rfloor$. This indicates $t_1 = c_1$, and since $t_1 = r_1m + c_1$, $r_1m = 0$. Since multiplier $m \geq 2$, $r_1 = 0$. Following this pattern, it can be noted that $r_0 = 1$ and $r_i = 0$ for all $i = 2, ..., l_r$. Therefore, $r$ can be represented as $[00...1]_b = [1]_b = 1$ for all instances of base $b$ and multiplier $m$. \vspace{\baselineskip}

\noindent
\textbf{Theorem 1:} For the path of carry values $c_i$'s that make the smallest closed loop across a multiplication transducer $T_{m,b}$ starting and ending from $0$, the read and write values are:
\begin{itemize}
  \item $r = [1]_b = 1$.
  \item $w = m$.
\end{itemize}

\noindent
\textit{Proof.} First, note that the smallest closed loop across a multiplication transducer $T_{m,b}$ must contain a $c_i \neq 0$. Therefore, in order to arrive at the smallest closed loop, the path needs to produce a non-zero carry value that will get the path back to the carry value of zero as quickly as possible. It needs to choose the smallest carries in order to arrive at this smallest path. \vspace{\baselineskip}

\noindent
Since the total is calculated as $t_0 = r_0m + c_0$ and $c_0 = 0$, the fastest way to get it to the smallest state/carry value is by making $r_0 = 1$. This is because $t_0 = r_0m$ and any $r_0 > 1$ would produce a state that is greater than what is produced by $r_0 = 1$ since the next state $c_1 = \lfloor{\frac{t_0}{b}}\rfloor = \lfloor{\frac{r_0m}{b}}\rfloor$.  Notice that this potrays a similar result to what was seen in the proof discussed in the first conjecture. \vspace{\baselineskip}

\noindent
If $r_0 = 1$, then $t_0 = m$. Therefore, the next carry value is $\lfloor{\frac{m}{b}}\rfloor$ while the read value is $w_i = t_0 ( \text{mod b}) = m (\text{mod b})$. Now that it has left state zero, it now aims to have the shortest path possible. In our expression of $t_i = r_im + c_i$, we can’t control anything except $r_i$ since $m$ is predefined and $c_i$ is dependent on the previous calculation. The simplest way to get the transducer back to state zero is to make $r_i = 0$. \vspace{\baselineskip}

\noindent
Therefore, $t_i = (0)(m) + c_i$, so $t_i = c_i$. Note that $t_1 = \lfloor{\frac{m}{b}}\rfloor$, and $c_2$ can be calculated by taking the integer division of the previous state and the base. $\square$ \vspace{\baselineskip}

\noindent
To understand this concept in further detail, let's take an example of base $b = 3$ and multiplier $m = 10$: \vspace{\baselineskip}

\noindent 
\textbf{Step 1} ($i=0$): $c_0 = 0$ (our initial state) and $r_0 = 1$. Therefore,
\begin{itemize}
  \item $t_0 = r_0m + c_0 = 1*10 + 0 = 10$
  \item $c_1 = \lfloor{\frac{t_0}{b}} \rfloor = \lfloor{\frac{10}{3}} \rfloor = 3$
  \item $w_0 = t_0 \text{ (mod b)} = 10 \text{ (mod 3) } = 1$
\end{itemize} \vspace{\baselineskip}

\noindent 
\textbf{Step 2} ($i=1$): $c_1 = 3$ and $r_1 = 0$. Therefore,
\begin{itemize}
  \item $t_1 = r_1m + c_1 = 0*10 + 3 = 3$
  \item $c_2 = \lfloor{\frac{t_1}{b}} \rfloor = \lfloor{\frac{3}{3}} \rfloor = 1$
  \item $w_1 = t_1 \text{ (mod b)} = 3 \text{ (mod 3) } = 0$
\end{itemize} \vspace{\baselineskip}

\noindent 
\textbf{Step 3} ($i=2$): $c_2 = 1$ and $r_2 = 0$. Therefore,
\begin{itemize}
  \item $t_2 = r_2m + c_2 = 0*10 + 1 = 1$
  \item $c_3 = \lfloor{\frac{t_2}{b}} \rfloor = \lfloor{\frac{1}{3}} \rfloor = 0$
  \item $w_2 = t_2 \text{ (mod b)} = 1 \text{ (mod 3) } = 1$
\end{itemize} \vspace{\baselineskip}

\noindent
In this example, notice that the read values follow a pattern similar to that shown in the proof of the first theorem. Essentially, the read value $r = [1]_b = 1$ for all bases. Additionally, an interesting pattern occurs when looking at the write values produced by the example above. The write value is $[101]_3 = [10]_{10}$. With a read value of $[1]_b$, the write values can always be generalized to have pattern $w = m$ as evidenced by the example above. Therefore, a general pattern for the read and write values of the shortest path have been found. \vspace{\baselineskip}

\noindent   
\textbf{Corollary 1:} The length for the path of carry values $c_i$'s, the smallest closed loop across a multiplication transducer $T_{m,b}$ starting and ending at $0$, is $\lfloor{m^{1/b}}\rfloor + 2$ for $b \geq 2$. \vspace{\baselineskip}

\noindent
\textit{Proof:} Note that this combination of read values always produces the shortest path as it goes to the state that is just far enough to escape zero, and then it takes the fastest approach to go back to zero afterwards. Knowing this, the length of the shortest path is $\lfloor{m^{1/b}}\rfloor + 2$ for $b \geq 2$ by using arithmetic logic. 

\vspace{\baselineskip}

\noindent
First, note that the addition of two is to account for the zeroes in the beginning and ending of the shortest path. For the path between these zeroes, $\lfloor{m^{1/b}}\rfloor$ can be used to denote the length. Remember that these numbers are $c_i = \lfloor{\frac{t_{i-1}}{b}}\rfloor$ where $t_0 = m$ and $t_i = c_i$. Knowing this is the case, that means that there is an integer division between m and $b^{l_w-1}$ at the very last step where $l_w$ represents the length of the write value. This is because base $m$ is divided by base $b$ in all steps of this base division except for the first since $t_i$ equals $c_i$. \vspace{\baselineskip}

\noindent
Knowing that the integer division needs to produce a zero in order to produce a closed path, then $\frac{m}{b^{l_w -1}} < 1$ or $m < b^{l_w - 1}$. Since $l_w - 1= l_p$ where $l_p$ represents the length of the smallest path of base $b$ and multiplier $m$ (excluding the first and last zeroes), this formula can be rewritten as $m < b^{l_p}$. Doing algebraic manipulation on this inequality, we can reasonably conclude that the length of the smallest closed set (excluding the zeroes) or $l_p$ of base $b$ and multiplier $m$ is $\lfloor{m^{1/b}}\rfloor$. Therefore, we can conclude that the length of the smallest closed set (including the zeroes) of a particular base $b$ and multiplier $m$ is $\lfloor{m^{1/b}}\rfloor + 2$.  $\square$ \vspace{\baselineskip}

\noindent
\textbf{Corollary 2:} The multipliers that have a length of $n + 1$ for the shortest closed loop across a multiplication transducer $T_{m,b}$ starting and ending at $0$ for a particular $b$ has a range of $m \in [b^{n-1}, b^n-1]$ for all $n \geq 3$ and $b \geq 2$. Therefore, the number of multipliers that have a length of $n + 1$ for a particular $b$ is $b^{n-1}(b-1)$ for all $n \geq 3$ and $b \geq 2$. \vspace{\baselineskip}

\noindent
\textit{Proof.} We prove this corollary by proving that there are sharp bounds for multipliers that have a length of $n + 1$ for the shortest closed loop and that the length of the shortest closed loop with respect to multipliers is monotonically increasing. \vspace{\baselineskip}

First, note that Theorem 1 shows that the length of the shortest closed loop with respect to multipliers in monotonically increasing, since $f(m) = \lfloor{m^{1/b}}\rfloor + 2$ is monotonically increasing.\vspace{\baselineskip}

We then prove the lower bound of $m = b^{n-1}$ is sharp. Let $m = b^{n-1}$. Then, we try to prove that the length of the shortest closed loop around a multiplication transducer $T_{b^{n-1}, b}$ starting and ending at $0$ for any $b$ is $n + 1$. Note that the first two elements in the path are $0$ and $\lfloor{\frac{b^{n-1}}{b}}\rfloor = \lfloor{b^{n-2}}\rfloor$. The third element in the path is $\lfloor{\frac{b^{n-2}}{b}}\rfloor = \lfloor{b^{n-3}}\rfloor$. This means that the ith element is $\lfloor{b^{n-i}}\rfloor$ and the nth element is $\lfloor{b^{n-n}}\rfloor = b^0 = 1$. Therefore, the (n+1)th element is $\lfloor{\frac{1}{m}}\rfloor = 0$. We can see that the length is $n + 1$ for $m = b^{n-1}$.  \vspace{\baselineskip}

\noindent
The lower bound can be shown to be $m = b^{n-1}$ by showing $m = b^{n-1} - 1$ has a length of $n$. Note the second element in the path would be $\lfloor{\frac{b^{n-1} - 1}{b}}\rfloor = \lfloor{b^{n-2} - 1 }\rfloor$. The third element in the path would be $\lfloor{\frac{b^{n-2} - 1}{b}}\rfloor = \lfloor{b^{n-3} - 1}\rfloor$.This means that the ith element is $\lfloor{b^{n-i}-1}\rfloor$ and the nth element is $\lfloor{b^{n-n}-1}\rfloor = b^0 - 1 = 1-1 = 0$. Therefore, the length is $n$ for $m = b^{n-1} - 1$. \vspace{\baselineskip}

\noindent
Next, the upper bound has to be shown to be $m = b^n -1$. We can do this by proving that $b^n$ has a length of $n + 2$. Note that the proof that the length of $m = b^{n-1}$ is $n + 1$ can be altered such that the length of $m = b^n$ is $n + 2$. We then show that $m = b^n-1$ has a length of $n + 1$. Note that the proof for the length of $m = b^{n-1} - 1$ is $n$ can be altered such that the length of $m = b^n -1$ has a length of $n + 1$. Therefore, the upper and lower bounds have been proven to be sharp and the lengths are monotonically increasing with respect to the multipliers. Thus, the corollary is proven. $\square$ \vspace{\baselineskip}

\section{Conclusion}

\hspace*{5mm} The theorems shown above provide an overview of multiplication transducers with no excluded digits and analyze paths through the multiplication transducer when $m = 1$. These help prove the basis of multiplication transducers inside of base multiplications and will help with rapidly calculating bases with small multipliers. \vspace{\baselineskip}

Further research includes generalizing the theorems to multiplication transducers with a reduction of the digit set and determining whether some of the same properties hold. Additionally, Corollary 1 has yet to be analytically proven, which proves another topic of exploration. Furthermore, pathways with larger multipliers can be explored, and better ways of visualizing multiplication transducers with large $b$'s and $m$'s have yet to be discovered.

\subsection{Exploring Quotient Sets With Restricted Digits}

\hspace*{5mm} As seen above, with these multiplication transducers, we can calculate an output $w$ when multiplying $m$ by $r$ in base $b$. We can now add a further constraint to limit the number of $r$ values that can be multiplied by $m$, which will, in turn, reduce the set of all outputs and reduce the number of states in the multiplication transducer. \vspace{\baselineskip}

This constraint involves reducing the original digit set $\{d_1, d_2, ..., d_k\}$, the set of digits that can be used to represent $r$ in base $b$, where $ k = b$. For instance, for digit set $\{0, 1\}$ for $b = 3$, $r = \{1, 3, 4, 9, ...\} = \{[1]_3, [10]_3, [11]_3, [100]_3, ...\}$, and for digit set $\{0, 1, 2\}$ (the entire set for base three), $r = \{1, 2, 3, 4, ...\} = \{[1]_3, [2]_3, [10]_3, [11]_3, ...\} = \mathbb{N}$. \vspace{\baselineskip}

Let $S (b; \{d_1, ..., d_k\})$ be the set of all $r$ that can be created in base $b$ using digit set $\{d_1, ..., d_k\}$. For the previous example, $S (3; \{0, 1\}) = \{1, 3, 4, 9, ...\} $. To express this mathematically,

\begin{equation} S (b; \{d_1, ..., d_k\}) = \{s \in \mathbb{N}; s = \sum_{i=0}^{\infty} \alpha_i b^i \text{ with } \alpha_i \in \{d_1, ..., d_k\} \text{ for all } i \} \end{equation}

What we are interested in doing is studying the positive whole numbers that come from dividing numbers in $S (b; \{d_1, ..., d_k\})$ or the set of all $q$ for a particular $b$. This set is known as a quotient set, and is denoted as $Q (b; \{d_1, ..., d_k\})$. Expressed mathematically,

\begin{equation} Q (b; \{d_1, ..., d_k\}) = \{x \in \mathbb{Z}: x = \frac{s}{s'} \text{ for some } s, s' \in S(b; \{d_1, ..., d_k\})\}\end{equation}

Note that in $Q (b; \{d_1, ..., d_k\})$, $s' \neq 0$. \vspace{\baselineskip}

We can prove whether a particular number $n$ is in this quotient set if two conditions are met: 

\begin{itemize}
  \item $w = n$.
  \item $p_0$ = $p_{l_{w}-1}$ = $0$ (or there is a closed loop in the multiplication transducer starting and ending at 0).
\end{itemize}

Note that for no restricted digits (containing the original digit sit), $Q = \mathbb{N}$.

\printbibliography
\end{document}